\documentclass[aps,showpacs,notitlepage,superscriptaddress,longbibliography]{revtex4-1}

\usepackage{epsfig,graphics,amssymb,amsmath,subeqnarray,color,bm,bbm}
\usepackage[toc,page]{appendix}
\usepackage{mathtools}
\usepackage{xcolor}
\usepackage{float}
\usepackage{graphicx}
\usepackage[colorlinks=true,allcolors=blue]{hyperref}
\usepackage{dsfont}
\usepackage{dutchcal}
\UseRawInputEncoding

\usepackage{mathtools}

\DeclarePairedDelimiterXPP\ev[1]{E}[]{}{

#1}
\begin{document}
\title{Long-range Velocity Correlations from Active Dopants}

\newcommand{\gaug}
{\affiliation{Institute for Theoretical Physics, University of G{\"o}ttingen, Friedrich-Hund-Platz 1, 37077 G\"ottingen, Germany}}

\newcommand{\mtl}
{\affiliation{Max Planck School Matter to Life, University of G{\"o}ttingen, Friedrich-Hund-Platz 1, 37077 G{\"o}ttingen, Germany.}}

\newcommand{\dcs}
{\affiliation{Institute for the Dynamics of Complex Systems, University of G{\"o}ttingen, Friedrich-Hund-Platz 1, 37077 G{\"o}ttingen, Germany.}}

\newcommand{\mpids}
{\affiliation{Max Planck Institute for Dynamics and Self-Organization, Am Fassberg 17, 37077 G{\"o}ttingen, Germany.}}

\author{Leila Abbaspour }

\dcs
\mtl
\mpids

\author{Rituparno Mandal}
\gaug

\author{Peter Sollich}
\gaug
\affiliation{Department of Mathematics, King's College London, London WC2R 2LS, UK}

\author{ Stefan Klumpp }

\dcs
\mtl

\def\red{\textcolor{red}}
\def\blue{\textcolor{blue}}


\begin{abstract}

One of the most remarkable observations in dense active matter systems is the appearance of long-range velocity correlations without any explicit aligning interaction (of e.g.\ Vicsek type). Here we show that this kind of long range velocity correlation can also be generated in a dense athermal passive system by the inclusion of a very small fraction of active Brownian particles. We develop a continuum theory to explain the emergence of velocity correlations generated via such active dopants. We validate the predictions for the effects of magnitude and persistence time of the active force and the area fractions of active or passive particles using extensive Brownian dynamics simulation of a canonical active-passive mixture. Our work decouples the roles that density and activity play in generating long range velocity correlations in such exotic non-equilibrium steady states.

\end{abstract}

\maketitle
\paragraph{Introduction:}

Active matter systems are one of the best-known examples of non-equilibrium systems and are famous for their fascinating collective behaviour across a diverse range of length and time scales~\cite{rmp1,rmp2}, from the cytoskeleton to bacterial colonies, tissues, flocks of birds to animal  herds. 

Systems of active Brownian particles (ABP), i.e.\ particles exhibiting self-propulsion, are a canonical example of active matter~\cite{marchetti12,takatori15,tailleur15,levis17,solon18}. These systems exhibit two types of non-equilibrium pattern formation: in the presence of aligning interactions between the directions of the self-propelled motion of the particles, they show flocking, i.e.\ collective directed motion of groups of such particles~\cite{mora2016,Ballerini2017}.  In the absence of such aligning interactions, they exhibit motility induced phase separation (MIPS), crucially without the need for attractive interactions~\cite{marchetti12,tailleur15,takatori15}. It is important to note that this phase separation is not associated with any macroscopic order in the orientation of the self-propulsion directions of the particles.
With external driving, on the other hand, such order can appear, see e.g.\ Ref.~\cite{mandal21}.

\begin{figure}
    \centering
   \includegraphics[width=\linewidth]{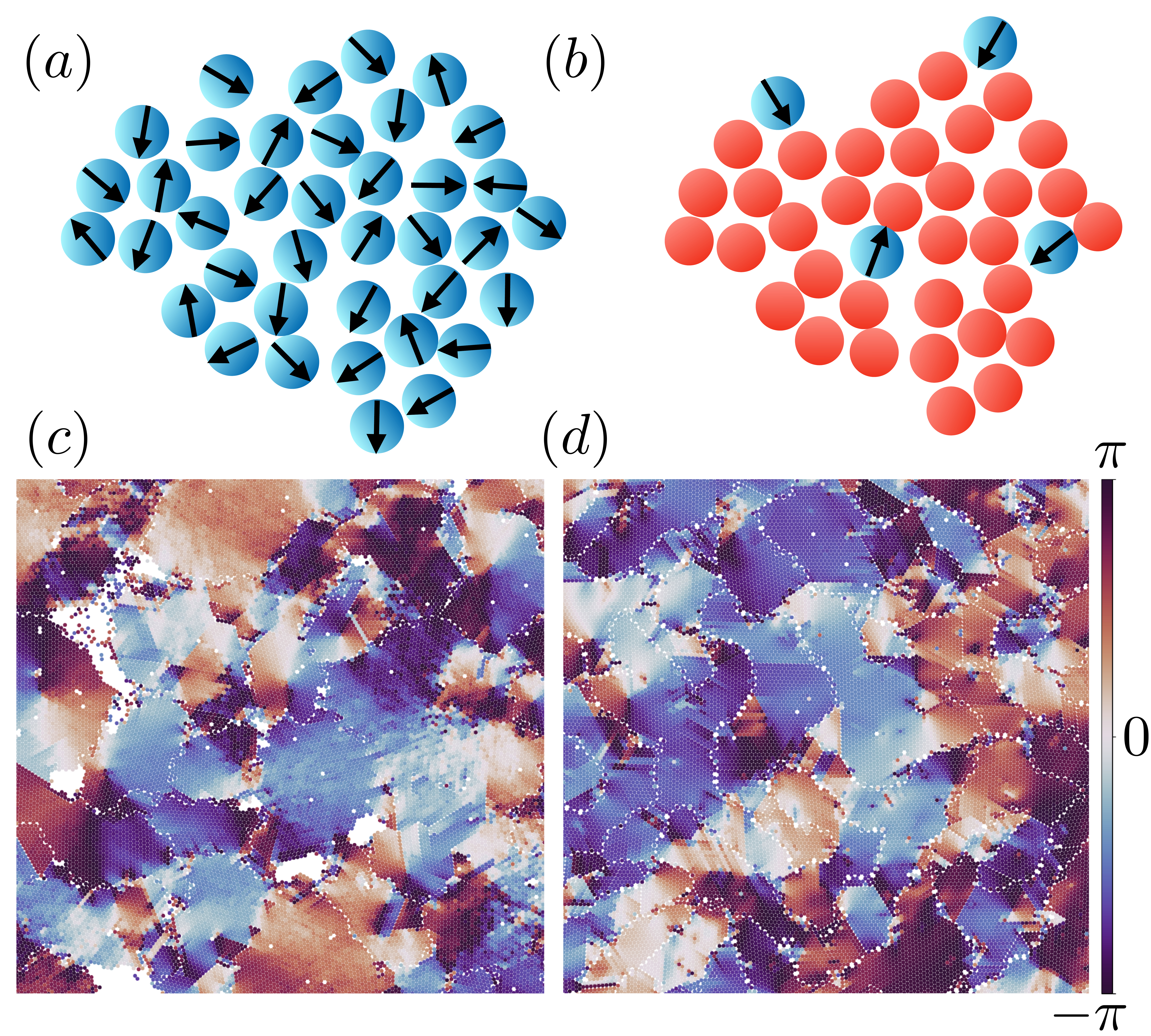}
    \caption{Schematic illustration of (a) a dense assembly of active particles, and (b) a dense mixture of active (blue) and passive (red) particles. Snapshots of the system with each particle coloured according to the orientation of its instantaneous velocity vector, for (c) $\phi_a=0.7$ ($\phi_p=0$), (d) $\phi_a=0.01$, $\phi_p=0.69$; both systems show strong emergent velocity correlations.
    }
    \label{Fig1}
\end{figure}

Very recently it has been discovered that while systems  without aligning interactions show no macroscopic orientational ordering, there are spectacularly large spatial structures in the instantaneous velocity field, especially within the dense phase created by motility-induced phase separation (MIPS)~\cite{Caprini2020-a,Caprini2020-b,Caprini2020-c,Henkes2020,Szamel2021}. It has been shown using both analytical calculations and numerical models (and confirmed in experiments in dense tissues~\cite{Henkes2020}) that in such a scenario, a dense assembly of active particles generates long range velocity correlations in the large persistence time limit; the corresponding correlation length grows as a power law $\sim \sqrt{\tau_p}$ with increasing persistence time $\tau_p$~\cite{Caprini2020-a,Caprini2020-b,Caprini2020-c,Henkes2020,Szamel2021}. The emergence of such non-equilibrium velocity correlation has always been attributed to a (highly persistent) dense active matter system and thus taken to require a high density of active particles. 

These results raise the question of whether the above two conditions, of high density and high activity, can be decoupled. In particular, could long-range velocity correlations be generated in a dense system of {\em passive} particles, by introducing activity only through a small fraction (e.g.\ much lower than the percolation density) of active particles?
A related question of interest is how ordered non-equilibrium states are affected by the inclusion of defect particles (e.g.\ static defects or motile non-aligning agents known as dissenters) that do not participate in the processes that induce the order. The role of both quenched~\cite{Yllanes2017, Rakesh2018, Martinez2018} and annealed~\cite{Yllanes2017, Martinez2018, Bera2020} disorder in the context of Vicsek-like models has been investigated very recently and it has been shown that presence of both types of disorder tends to destroy the ordered flocking state~\cite{Yllanes2017, Rakesh2018, Martinez2018,Bera2020}. Our work explores a similar line of questions, but in a system where long-range velocity correlations appear without any explicit alignment interaction. We ask in particular whether the long-range order in the instantaneous velocity field seen in such systems is stabilised or destabilized by the inclusion of a large fraction of passive particles. 

Motivated by the above questions, in this paper we study mixtures of active and passive particles to explore whether inclusion of passive particles enhances or suppresses local orientational order and to see whether we can decouple the roles of activity and density in generating long-range velocity correlations. Using extensive particle-based simulation of an active-passive mixture, we demonstrate that velocity ordering is enhanced by an increasing density of passive particles, and show that long-range velocity correlations can be generated in an athermal passive medium by a tiny fraction of active insertions (dopants) as long as the medium is dense enough. We also construct an analytical theory to explain the physics of velocity correlations in a dense passive medium with active dopants. Our hydrodynamic theory predicts that the amplitude of the velocity correlations is proportional to $f^2$ where $f$ is the magnitude of the propulsion force acting on each active particle, proportional to $\tau_p^{-3/4}$ where $\tau_p$ is the persistence time of the active particles, and proportional to the density of active particles $\phi_a$. The hydrodynamic theory also predicts that the correlation length $\xi_L$ only depends on $\tau_p$, as $\sqrt{\tau_p}$~\cite{Caprini2020-a,Caprini2020-b,Caprini2020-c,Henkes2020,Szamel2021}, not on the active forcing magnitude $f$. We verify these theoretical predictions by performing further targeted simulations. The explicit form of the correlation function that we derive theoretically decouples the roles that density and activity play in generating long range velocity correlations in a non-equilibrium steady state. This insight will be useful in understanding long range ordering in e.g.\ dense passive colloidal systems driven by a few self-propelled Janus colloids, or assembly of dead bacteria churned up by few living ones.

\paragraph{{Particle-based Model:}}

We consider a binary mixture~\cite{Leila2021} of passive and active Brownian particles (ABP)~\cite{marchetti12,takatori15,levis17,solon18,tailleur15} moving in two dimensions and occupying  area fractions of $\phi_p$ and  $\phi_a$, respectively. The dynamical evolution of the particle positions is described by the overdamped equations of motion:

\begin{equation}
    \gamma  \dot{ \mathbf{r}}_i = \mathbf{F}_i + f\mathbf{n}_i\,  \Delta_{i}(\mathcal{A})
    \label{eom}
\end{equation}
where $\mathbf{r}_i$ is the position vector of the $i$-th particle and $\gamma$ is the constant drag coefficient governing the friction force acting on each particle. The factor $\Delta_i(\mathcal{A})$ ($\Delta_i=1$ for the active particles and $\Delta_i=0$ for the passive particles)  restricts the active forces $f \mathbf{n}_i$ to the particles in the subset $\mathbcal{A}$ of active particles. The orientation vectors of the active forces are  
$\mathbf{n}_i=(\cos{\theta_i},\sin{\theta_i})$, and the orientation angles 
$\theta_i$ of the active forcing follow the dynamics 
\begin{equation}
    \mathbf{\dot \theta}_i = \sqrt{\frac{2}{\tau_p}}\zeta_i \quad \text{for} \quad i \in \mathbcal{A}
    \label{eofth}
\end{equation}
where the noise $\zeta_i$ has zero mean and time correlations  $\langle\zeta_i (t) \zeta_j (t^\prime)\rangle =\delta_{ij} \delta (t-t^\prime).$ In Eq.~\ref{eofth},  $\tau_p$ is the persistence time. All particles in the system interact only through steric interactions described by the forces $\mathbf{F}_i=-\nabla_i U$ where $U$ is a repulsive WCA (Weeks-Chandler-Anderson)  interaction potential~\cite{weeks1971} (see supplementary information for the details of the potential).

\begin{figure} 
    \centering
   \includegraphics[width=\linewidth]{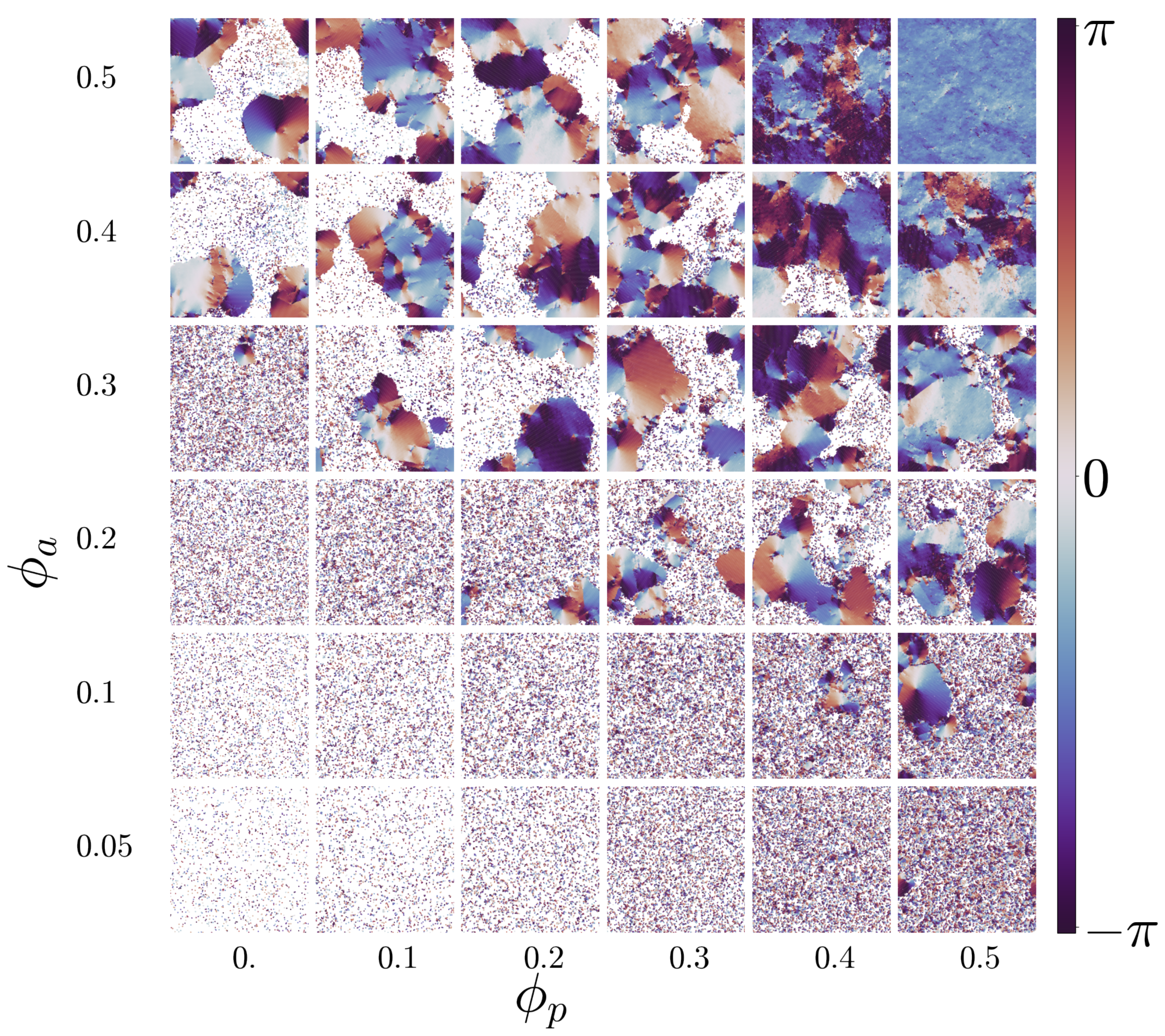}
    \caption{Snapshot of the system (binary mixture of active and passive particles) for various combinations of values of the area fractions of active and passive particles, $\phi_a$ and $\phi_p$, respectively. The snapshots at the bottom right corner shows that a small fraction of active inclusions can cause long-range velocity correlations in a dense athermal and almost entirely passive system.
    }
    \label{Fig-2}
\end{figure}

\paragraph{Velocity correlations in active--passive mixture:}

First we reproduce the long range velocity correlations in a completely active system (see Fig.\ref{Fig1}(a) for a schematic and Fig.\ref{Fig1}(b) for a snapshot of the system showing the long range ordering of the instantaneous velocities). This effect has been reported before in the context of different \emph{dense active matter} systems, both in simulations and in experiments~\cite{Caprini2020-a,Caprini2020-b,Caprini2020-c,Henkes2020,Szamel2021}. Here we want to explore a different scenario (see Fig.\ref{Fig1}(c) for a schematic) and test whether a dense \emph{passive} system that is driven by just a few active particles can show similar ordering. Indeed, Fig.\ref{Fig1}(d) shows remarkably similar order in such a dense passive system that is driven by a very small fraction ($\phi_a=0.01$) of active Brownian particles. To explore this further, we also ran simulations for combinations of different fractions of active and passive particles (see Fig.~\ref{Fig-2} for snapshots). Strong local velocity correlations are seen to emerge as long as the total density $\phi_{\rm tot}=\phi_a+\phi_p$ is high enough and there exists a non-zero fraction of active particles. To get a better understanding of this velocity ordering in an active-passive mixture we developed a hydrodynamic theory that we describe in the next paragraph.

\begin{figure}
    \centering
   \includegraphics[width=\linewidth]{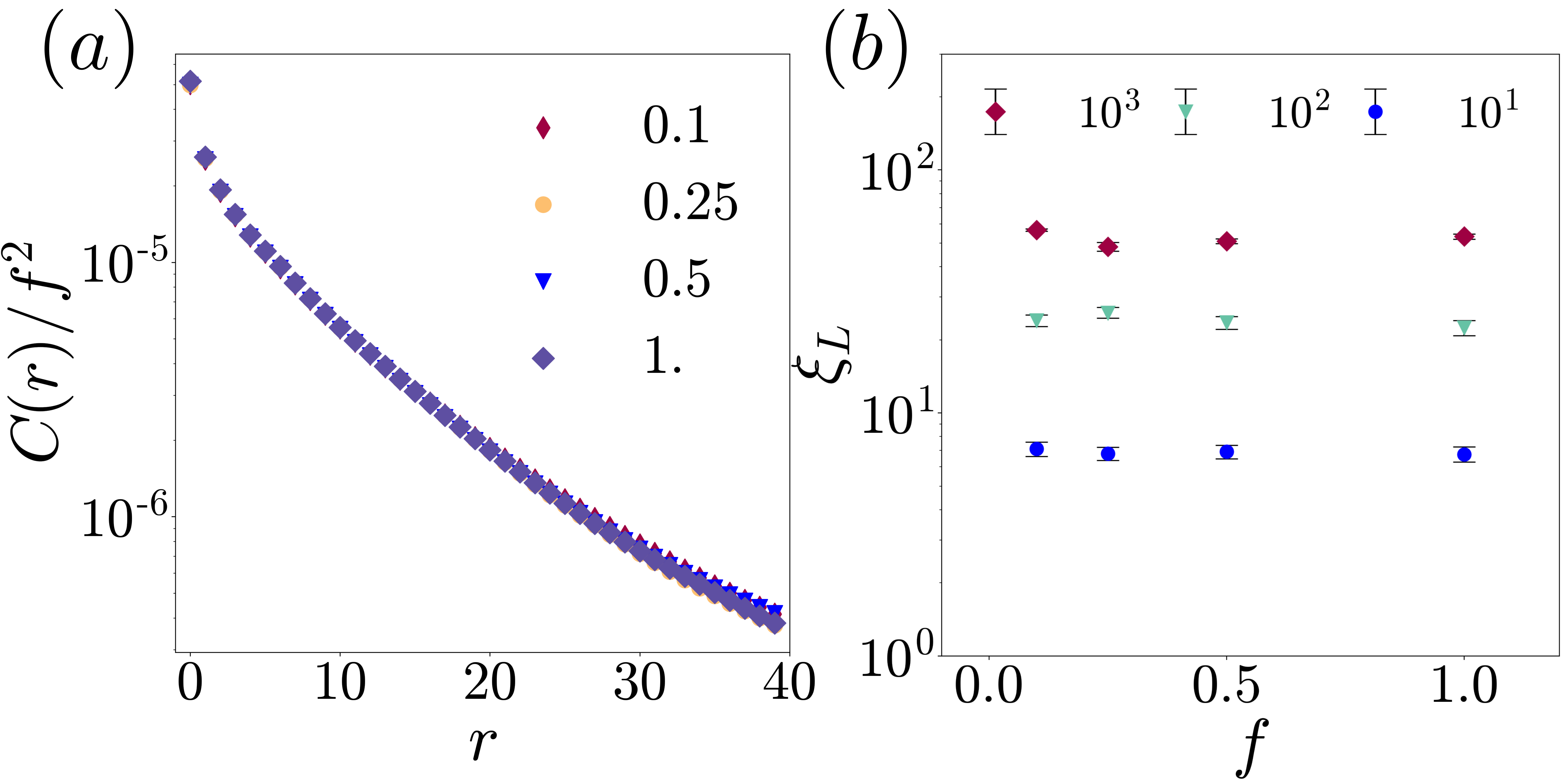}
    \caption{(a) Scaled velocity-velocity spatial auto correlation as a function of spatial distance $r$, for different active force magnitudes $f$ as indicated in the legend, and $\tau_p=3$ (b) The correlation length $\xi_L$ as a function of the active force for different persistence times $\tau_p$ as shown; the plot demonstrates that, at fixed persistence time, the correlation length is independent of the magnitude of the active force.
    }
    \label{Fig-3}
\end{figure}



\paragraph{Theory:}

To derive the correlation of the hydrodynamic velocity field of the system, we extend the approach used by Henkes {\it{et al.}}\ for a dense active system \cite{Henkes2020} to the case of a passive system with active dopants. We consider the entire collection of passive particles as a dense medium having a smooth velocity field, with the active particles as random point-like defects that are the sources of a force density of the form
\begin{equation}
 \mathbf{F}_i(\mathbf{r},t)=   a^2 f \mathbf{n}_i(t) \delta(\mathbf{r}-\mathbf{r}_i(t))
\end{equation}
for the $i$-th active particle. Here $a$ is a microscopic length scale given by the particle size, $f$ is the magnitude of the propulsion force acting on each active particle as before, and $\textbf{n}_{i}$ is the unit vector associated with the orientation of this active force.
Assuming now that in the large persistence time limit and at sufficiently high density, the active particles deform the elastic solid-like medium by pushing other particles without changing their positions significantly, we can evaluate the correlation between the forces from active particles $i$ and $j$ in Fourier space ($\mathbf{q},\omega$) as 
\begin{equation}\langle
\mathbf{F}_i(\mathbf{q},\omega) \cdot \mathbf{F}_j(\mathbf{q^\prime},\omega^\prime)\rangle 
 = \frac{4 \pi  \tau_p a^4 {f}^2}{1+(\tau_p \omega)^2}    e^{i \mathbf{q}\cdot \mathbf{r}_i} e^{i \mathbf{q}^\prime\cdot \mathbf{r}_j} \delta_{ij}
  \delta (\omega+\omega^\prime)  
\end{equation}
where the frequency dependence arises from the dynamics of the active force orientation $\mathbf{n}_i(t)$~\cite{Henkes2020}.
We then sum over all the particles (over the indices $i$ and $j$) and also over the steady state probability distribution of the active particles' positions $\mathbf{r}_i$, which we take as uniform, to arrive at the total force correlator
\begin{equation}
\langle \mathbf{F}(\mathbf{q},\omega) \cdot \mathbf{F}(\mathbf{q}^\prime,\omega^\prime)\rangle =\frac{N_a a^2 f^2 (2 \pi)^3}{N}   \delta(\mathbf{q}+\mathbf{q^\prime}) \frac{2\tau_p}{1+(\tau_p \omega)^2}
\end{equation}
where $N_a$ is the number of active particles and $N$ the total number of particles in the system. After taking an angular average, the form of the velocity correlation function can be derived~\cite{Henkes2020} (see supplementary information for the full calculation and the final form of the correlation function). Its behaviour can be approximated at large distances  as
\begin{equation}
    C(\mathbf{r}) \approx \left(\frac{\phi_a}{\phi_a+ \phi_p} \right) \left(\frac{a^2 {f}^2}{4 \pi \zeta^2} \sqrt{\frac{\pi}{2 r}}\right) \frac{1}{{\xi_L}^{3/2}}e^{-{r/\xi_L}}
    \label{theory}
\end{equation}
where $\xi_L=\left(\frac{(B+\mu)\tau_p}{\zeta} \right)^{1/2}$, $r$ is the spatial distance, $\zeta$ is the friction coefficient as before, and $B$ and $\mu$ are the bulk and shear moduli of the overall medium. These are expected to be dependent only on the total area fraction 
$\phi_{\rm tot}=\phi_a+\phi_p$ of active and passive particles. Therefore we now have a testable prediction about the equal time velocity auto correlation function from our hydrodynamic theory in terms of the control parameters $f, \tau_p, \phi_a, \phi_p$ etc.

\begin{figure}
    \centering
   \includegraphics[width=\linewidth]{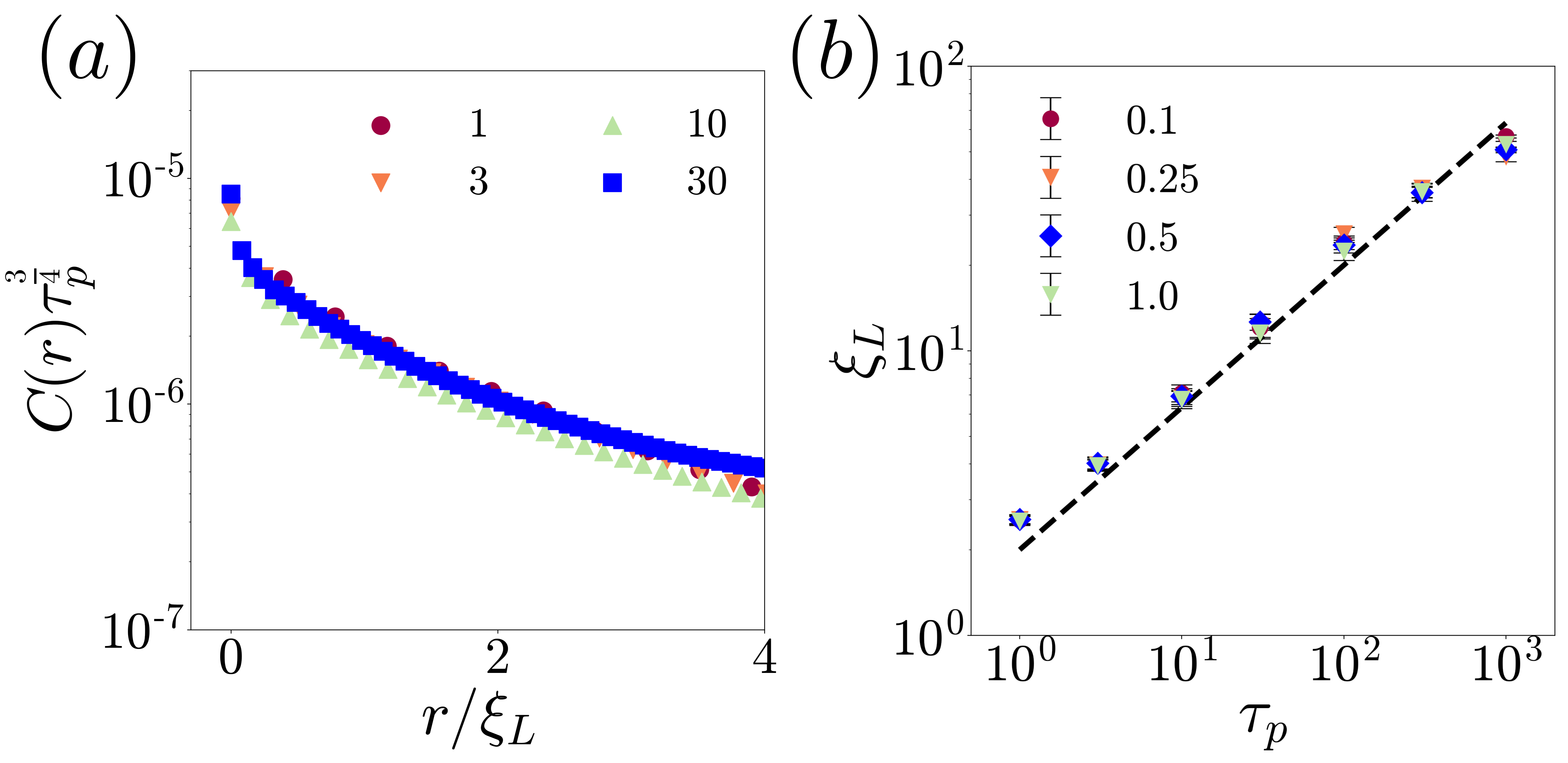}
    \caption{(a) The scaled equal-time velocity-velocity correlation as a function of spatial distance $r$ for different persistence times $\tau_p$ as indicated, for active force magnitude $f=0.25$. (b) The correlation length $\xi_L$ as a function of the persistence time $\tau_p$ for different values of the active force magnitude $f$ shows that the correlation length grows as a power law (with exponent $\frac{1}{2}$) with the persistence time $\tau_p$ and this behaviour is independent of the magnitude of the active force $f$, which is shown in the legend.
    }
    \label{Fig-4}
\end{figure}

\paragraph{Comparison with Theory:}
To validate the predictions of our theory we ran further simulations to test explicitly the effects of varying active force magnitude $f$, persistence time $\tau_p$, and area fractions of active $\phi_a$ and passive particles $\phi_p$, and compared the correlation functions from those simulations with those 
calculated using the hydrodynamic theory.

Eq.~(\ref{theory}) predicts that the prefactor of the velocity-velocity spatial auto-correlation function will scale as $f^2$ when we vary the active force magnitude $f$, without any associated variation in the correlation length. Therefore the values of the scaled auto-correlation function $C(r)/f^2$ are expected to collapse into a single curve for different values of the active forcing $f$ as long as the other parameters are kept constant. Fig.~\ref{Fig-3}(a) clearly shows that the scaled correlation functions for different magnitudes of the active force do indeed collapse on top of each other. In Fig.~\ref{Fig-3}(b) we show the correlation length ($\xi_L$) as a function of the active force $f$ for different persistence times ($\tau_p$), which provides further evidence that the correlation length is independent of the magnitude of the active force when the persistence time is kept constant.

To shed light on the effects of the persistence time  scale we vary $\tau_p$ next, keeping the both area fractions ($\phi_a, \phi_p$) and the active forcing magnitude $f$ constant. Our hydrodynamic theory (see Eq.~\ref{theory}) suggests that the prefactor of the velocity-velocity spatial auto-correlation function scales as $\tau_p^{-3/4}$ (due to the dependence of $\xi_L$ on $\tau_p$),  apart from the standard exponential dependence on $r/\xi_L$. Indeed as Fig.~\ref{Fig-4}(a) shows the simulation data points for different persistence times $\tau_p$ for a given active force $f$ fall nicely on the same curve once we scale $C(r)$ appropriately, i.e.\ $C$ by $\tau_p^{-3/4}$ and $r$ by $\xi_L$. Fig.~\ref{Fig-4}(b) further indicates that the correlation length $\xi_L$ grows as the square root of persistence time $\tau_p$ regardless of the magnitude of the active force. This is also consistent with our theory and the earlier studies~\cite{Caprini2020-a,Caprini2020-b,Caprini2020-c,Henkes2020,Szamel2021}.

\begin{figure}
    \centering
   \includegraphics[width=\linewidth]{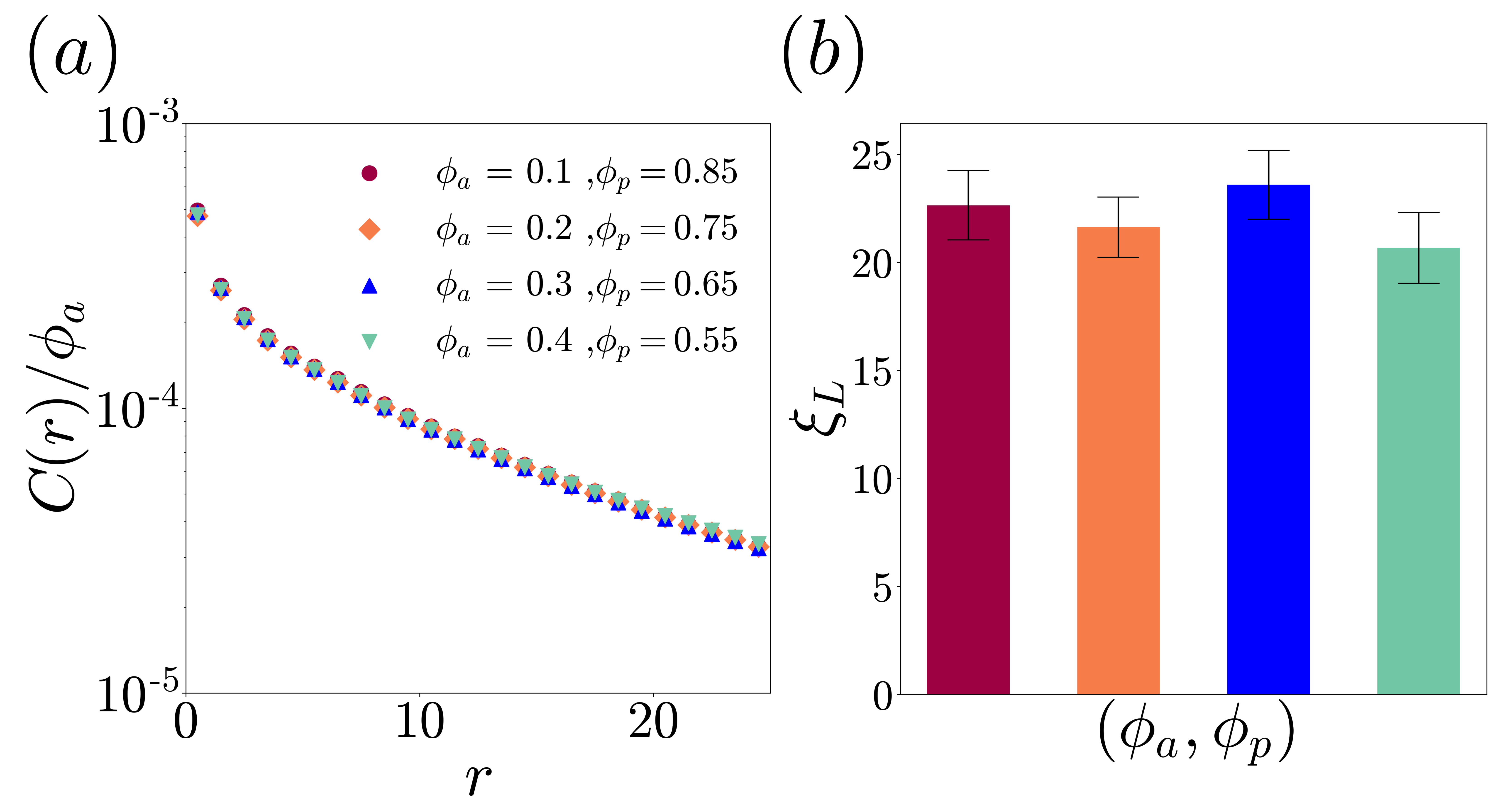}
    \caption{(a) The scaled velocity-velocity spatial auto correlation function as a function of spatial distance $r$ for different combinations of $\phi_a$ and $\phi_p$ for a fixed value of $\phi_{\rm tot}=0.95$. (b) Data for the correlation length are consistent 
    with $\xi_L$ being constant across 
    the same combinations of $\phi_a$ and $\phi_p$; the color scheme is the same as in (a).}
    \label{Fig-5}
\end{figure}

We finally explore the dependence on the area fractions of passive and active particles $\phi_p$ and $\phi_a$, respectively, while keeping the active forcing parameters ($f,\tau_p$) constant. The theory suggests that apart from the linear dependence on the fraction of active particles $\phi_a$ there is no separate dependence on $\phi_a$, i.e.\ all other density dependences of the correlation function $C(r)$ appear only via the total area fraction $\phi_{\rm tot}=\phi_a+\phi_p$ of the binary mixture. Our simulations, which involve different mixture compositions, confirm this prediction in Fig.~\ref{Fig-5} (a) where we scale the correlation function by the area fraction of active particles $\phi_a$ for a fixed value of the total density $\phi_{\rm tot}$. Fig.~\ref{Fig-5} (b) shows that the system has practically the same correlation length regardless of the fraction $\phi_a$ of active particles (or the fraction $\phi_p$ of passive particles) when the total area fraction $\phi_{\rm tot}$ is sufficiently high and kept constant.

\paragraph{Conclusion:}
In this article we have demonstrated that long-range velocity correlations (which have only been observed in dense active mater system until now ~\cite{Caprini2020-a,Caprini2020-b,Caprini2020-c,Henkes2020,Szamel2021}) can be generated in a dense athermal passive system by including a very small fraction of persistent active Brownian particles. This observation conceptually decouples the roles  played by density and activity parameters in generating such non-equilibrium ordering effects. Also, our results extend the discussion on whether inclusion of disorder can increase order in a  system or whether conversely it tends to destroy order ~\cite{Yllanes2017, Rakesh2018, Martinez2018,Bera2020}. 

We started by providing evidence that with a very small amount of active inclusions or dopants, an otherwise passive, dense athermal system can exhibit long range velocity correlations similar to a pure dense assembly of active particles. We explored the degree of velocity correlation for different  numbers of active and passive particles particles in such a mixture. We then derived the hydrodynamic theory to calculate the equal time velocity auto-correlation function in terms of the microscopic system parameters such as $f$, $\tau_p$ etc. This theory made testable predictions that we confirmed via further molecular dynamics simulations. 
We examined the impact of different parameters on the velocity correlations and found good agreement between the simulation results and the hydrodynamic theory. Specifically, we found that the correlation length depends only on the overall area fraction of active and passive particles and grows as $\sqrt{\tau_p}$ with the persistence time of self-propulsion. The latter result is in agreement with previous findings on purely active systems ~\cite{Caprini2020-a,Caprini2020-b,Caprini2020-c,Henkes2020,Szamel2021}.

Our predictions and results can be further tested both in simulations~\cite{sturmer2019,Leila2021,mandal2022} and in experiments, e.g.\ on mixtures of microbes and passive colloidal particles~\cite{williams2022}, assemblies of active and passive colloids~\cite{singh2017,mu2022}, mixture of mobile and immobile bacteria~\cite{patteson2018}, or active granular mixtures~\cite{kumar2014,sriram2022}. Understanding the decoupling of density and activity makes it possible not only to reproduce a long-range velocity correlation in different active-passive mixtures or assemblies but also paves the way for designing and controlling active matter for practical purposes, e.g.\ in the context of transport and mixing.



\section*{Acknowledgement}

This research was conducted within the Max Planck School Matter to Life, supported by the German Federal Ministry of Education and Research (BMBF) in collaboration with the Max Planck Society. The simulations were run on the GoeGrid cluster at the University of G\"ottingen, which is supported by the DFG (grant INST 186/1353-1 FUGG) and MWK Niedersachsen (grant no.\ 45-10-19-F-02). This project has received funding from the European Union’s Horizon 2020 research and innovation programme under Marie Sk\l odowska-Curie grant agreement no.\ 893128.

\bibliography{reference}

\pagebreak
\clearpage
\widetext
\begin{center}
\textbf{\large Supplemental Materials:\\ Long-range Velocity Correlation from Active Dopants}

\end{center}
\setcounter{equation}{0}
\setcounter{figure}{0}
\setcounter{table}{0}
\setcounter{page}{1}
\makeatletter
\renewcommand{\theequation}{S\arabic{equation}}
\renewcommand{\thefigure}{S\arabic{figure}}
\renewcommand{\bibnumfmt}[1]{[S#1]}
\renewcommand{\citenumfont}[1]{S#1}

\section{Hydrodynamic Theory}

We follow the continuum approach used by Henkes {\it{et al.}}~\cite{Henkes2020} but with a crucial difference in terms of describing the medium, which will be explained in detail below. We start by writing the equation of motion of the displacement field,
\begin{equation}\label{eq:eom}
    \zeta \dot{\mathbf{u}}(\mathbf{r})= \mathbf{F}(\mathbf{r},t) + \nabla \cdot \hat{\sigma}
\end{equation}
where $\mathbf{u}(\mathbf{r})$ is the displacement field of the whole medium (where we do not distinguish whether the medium is made of active or passive particles), $\zeta$ is the friction coefficient and $\hat{\sigma}$ is the stress tensor. We also define $\mathbf{F}(\mathbf{r},t)=\sum_i \mathbf{F}_i(\mathbf{r},t)$ where $\mathbf{F}_i(\mathbf{r},t)$ is the force due to the $i$-th active particle which is of the form,
\begin{equation}
 \mathbf{F}_i(\textbf{r},t)=   a^2 f \:\mathbf{n}_i(t)\: \delta(\mathbf{r}-\mathbf{r}_i(t))
\end{equation}
 where $a$ is a microscopic length scale typically given by the particle parameter, $f$ is the magnitude of propulsion force acting on each active particle, $\textbf{n}_{i}$ is the unit vector associated with the orientation of the active force of the $i$-th active particle and $\textbf{r}_i$ is the position of the $i$-th active particle. 
 
  We then write down the Fourier transform (both in space and time) of the active forces as,
\begin{equation}
\mathbf{F}_i(\mathbf{q},\omega)=a^2 f \int \int \mathbf{n}_i(t)\: \delta(\mathbf{r}-\mathbf{r}_i(t))\: e^{i \omega t} e^{i \mathbf{q}\cdot \mathbf{r}} dt\:\: d^2\mathbf{r}.
\end{equation}
 
 Next, considering a spatially static active particle configuration (described by $\mathbf{r}_i(t)\equiv \mathbf{r}_i$) where the orientation of the active particles can still evolve, we get
\begin{equation}
\mathbf{F}_i(\mathbf{q},\omega)=a^2 f \:e^{i \mathbf{q}\cdot \mathbf{r}_i}  \int \mathbf{n}_i(t)\: e^{i \omega t}\: dt\ .
\end{equation}
We can now calculate the force correlation as,
\begin{equation}
 \mathbf{F}_i(\mathbf{q},\omega) \cdot \mathbf{F}_j(\mathbf{q^\prime},\omega^\prime) = a^4 {f}^2\:e^{i \mathbf{q}\cdot \mathbf{r}_i}\:e^{i \mathbf{q^\prime}\cdot \mathbf{r}_j}  \int \int  \mathbf{n}_i(t)\cdot \mathbf{n}_j(t^\prime) e^{i\omega t}\:e^{i\omega^\prime t^\prime} dt\: dt^\prime
\end{equation}
and with an appropriate average over an ensemble of configurations of force orientations we get,
\begin{align*}
 \langle \mathbf{F}_i(\mathbf{q},\omega) \cdot \mathbf{F}_j(\mathbf{q^\prime},\omega^\prime) \rangle &= a^4 {f}^2\:e^{i \mathbf{q}\cdot \mathbf{r}_i}\:e^{i \mathbf{q^\prime}\cdot \mathbf{r}_j}  \int \int \langle \mathbf{n}_i(t)\cdot \mathbf{n}_j(t^\prime) \rangle e^{i\omega t}\:e^{i\omega^\prime t^\prime} dt\: dt^\prime \\
 &=a^4 {f}^2\:e^{i \mathbf{q}\cdot \mathbf{r}_i}\:e^{i \mathbf{q^\prime}\cdot \mathbf{r}_j}  \int \int e^{-\frac{\vert t-t^\prime \vert}{\tau_p}}\:\delta_{ij} e^{i\omega t}\:e^{i\omega^\prime t^\prime} dt\: dt^\prime\\
 &= a^4 {f}^2 \:\delta_{ij} \:e^{i \mathbf{q}\cdot \mathbf{r}_i}\:e^{i \mathbf{q^\prime}\cdot \mathbf{r}_j} \: 2\pi \:\frac{2\tau_p}{1+(\tau_p \omega)^2}\: \delta (\omega+\omega^\prime)   
\end{align*}

Note that this correlation still depends on a specific configuration in terms of the positions of the active particles $\{ \mathbf{r}_i \}$. Summing over the particles
 \begin{align*}
  \sum_{ij} \langle \mathbf{F}_i(\mathbf{q},\omega) \cdot \mathbf{F}_j(\mathbf{q^\prime},\omega^\prime) \rangle &= \sum_{ij} a^4 {f}^2 \:\delta_{ij} \:e^{i \mathbf{q}\cdot \mathbf{r}_i}\:e^{i \mathbf{q^\prime}\cdot \mathbf{r}_j} \: 2\pi \:\frac{2\tau_p}{1+(\tau_p \omega)^2}\: \delta (\omega+\omega^\prime) \\&= \sum_{i} a^4 {f}^2 \:e^{i (\mathbf{q}+\mathbf{q^\prime}) \cdot \mathbf{r}_i} \: 2\pi \:\frac{2\tau_p}{1+(\tau_p \omega)^2}\: \delta (\omega+\omega^\prime),
 \end{align*}
and averaging over their positions gives the force correlation
  \begin{align*}
  \langle \mathbf{F}(\mathbf{q},\omega) \cdot \mathbf{F}(\mathbf{q^\prime},\omega^\prime) \rangle &= a^4  \sum_{\mathbf{r}_i} P(\mathbf{r}_i) {f}^2 \:e^{i (\mathbf{q}+\mathbf{q^\prime}) \cdot \mathbf{r}_i} \: 2\pi \:\frac{2\tau_p}{1+(\tau_p \omega)^2}\: \delta (\omega+\omega^\prime) \\ &= N_a a^4  \sum_{\mathbf{r}} P(\mathbf{r}){f}^2 \:e^{i (\mathbf{q}+\mathbf{q^\prime}) \cdot \mathbf{r}} \: 2\pi \:\frac{2\tau_p}{1+(\tau_p \omega)^2}\: \delta (\omega+\omega^\prime) \\ &=\frac{N_a {f}^2 a^4 }{L^2}  \int d^2\mathbf{r}  \:e^{i (\mathbf{q}+\mathbf{q^\prime}) \cdot \mathbf{r}} \: 2\pi \:\frac{2\tau_p}{1+(\tau_p \omega)^2}\: \delta (\omega+\omega^\prime) \\ &=\frac{N_a {f}^2 a^4 (2 \pi)^3}{L^2}  \delta(\mathbf{q}+\mathbf{q^\prime}) \:\frac{2\tau_p}{1+(\tau_p \omega)^2}\: \delta (\omega+\omega^\prime) \\ &=\frac{N_a  {f}^2 (2 \pi)^3}{N} \frac{a^4 N}{L^2} \delta(\mathbf{q}+\mathbf{q^\prime}) \:\frac{2\tau_p}{1+(\tau_p \omega)^2}\: \delta (\omega+\omega^\prime)\\ &=\frac{N_a}{N} a^2  {f}^2 (2 \pi)^3  \delta(\mathbf{q}+\mathbf{q^\prime}) \:\frac{2\tau_p}{1+(\tau_p \omega)^2}\: \delta (\omega+\omega^\prime)  
 \end{align*}
 where $N_a$ is the number of active particles, $N$ is the total number of particles and $L$ is the linear dimension of the system. This finally gives us the force correlation in the following form  
 \begin{equation}
     \langle \mathbf{F}(\mathbf{q},\omega) \cdot \mathbf{F}(\mathbf{q^\prime},\omega^\prime) \rangle=\frac{\phi_a}{\phi_a+ \phi_p} a^2  {f}^2 (2 \pi)^3  \delta(\mathbf{q}+\mathbf{q^\prime}) \:\frac{2\tau_p}{1+(\tau_p \omega)^2}\: \delta (\omega+\omega^\prime) 
     \label{forcecorr}
 \end{equation}
where $\phi_a, \phi_p$ are area fractions of active particles and passive particles respectively. Apart from the factor $\frac{\phi_a}{\phi_a+ \phi_p}$, this equation has the same form as derived in Ref.~\cite{Henkes2020} for a purely active system. 

The force correlation being known, we can follow Ref.~\cite{Henkes2020} to obtain the velocity correlations: taking the Fourier transform of the equation of motion \ref{eq:eom} and decomposing it into transversal and longitudinal modes, one can write the velocity correlation (in Fourier space) in terms of the force correlation (also in Fourier space) that we derived in Eq.~\ref{forcecorr}. Taking the inverse Fourier transform of the velocity correlation, we obtain
\begin{equation}
\begin{split}
    C_{vv}(\mathbf{r}) \approx \left(\frac{\phi_a}{\phi_a+ \phi_p} \right) \left(\frac{a^2 {f}^2}{4 \pi \zeta^2} \sqrt{\frac{\pi}{2 r}}\right) \left(\frac{1}{{\xi_L}^{3/2}}e^{-{r/\xi_L}}+\frac{1}{{\xi_T}^{3/2}}e^{-{r/\xi_T}} \right)
\end{split}
\end{equation}
where $r$ is the spatial distance. Here the correlation lengths are defined as $\xi_T=\left(\frac{\mu\tau_p}{\zeta} \right)^{1/2}$ and $\xi_L=\left(\frac{(B+\mu)\tau_p}{\zeta} \right)^{1/2}$ where $B$ and $\mu$ are the bulk and the shear modulus of the system \cite{Henkes2020}, which parameterize the stress tensor in Eq.~\ref{eq:eom}. If we make the further assumption that $B\gg\mu$ (which is most accurate for area fractions close to the jamming transition) this leads to the final expression for the velocity correlation function
\begin{equation}
    C(\mathbf{r}) \approx \left(\frac{\phi_a}{\phi_a+ \phi_p} \right) \left(\frac{a^2 {f}^2}{4 \pi \zeta^2} \sqrt{\frac{\pi}{2 r}}\right) \frac{1}{{\xi_L}^{3/2}}e^{-{r/\xi_L}}.
    \label{final}
\end{equation}
Eq.~\ref{final} shows the dependence on density ($\phi_a$, $\phi_p$) and activity parameters ($f$, $\tau_p$) explicitly in a closed-form expression.

\section{Details of the Particle Based Model}
The overdamped Brownian dynamics simulations were performed in two spatial dimensions. The equations of motion in Eq.~\ref{eom} were integrated with an explicit Euler-Maruyama scheme ~\cite{higham2001} with a time step $\Delta t=7 \times10^{-4}$ in a square box 
with periodic boundary conditions.
All the particles in the system interact only through the steric interactions where the force is defined as, $\mathbf{F}_i=-\nabla_i U_{tot}$. $U_{tot}$, the purely repulsive interaction potential of WCA type ~\cite{weeks1971} is given by
\begin{equation}
    U_{tot}(r_{ij})= 
\begin{cases}
     4 \epsilon \left[ \left(\frac{\sigma}{r_{ij}}\right)^{12} - \left(\frac{\sigma}{r_{ij}}\right)^6 \right] + \epsilon,& r_{ij} < 2^{1/6}\sigma\\
    0,              & \text{otherwise}
\end{cases}
\end{equation}
where $r_{ij}= |\mathbf{r}_i - \mathbf{r}_j|$ is the magnitude of separation vector of the pair of particles $i$ and $j$ and $\mathbf{r}_i$, $\mathbf{r}_j$ are the position vectors of the same pair. Here $\sigma$ stands for the parameter representing particle diameter and $\epsilon$ is the energy scale of the interactions. By using $\sigma=1.0$ and $\epsilon=1.0$ we set the scales of length and energy in our description to $\sigma$ and $\epsilon$, respectively. We choose a fixed box size of $L=107\sigma$ and various values of the area fraction $\phi_{\rm tot}$. The number of particles ($N$) in the system is then determined by $N=\frac{\phi_{\rm tot} 4L^2}{\pi\sigma^2}$. 
The number of active and passive particles ($N_a$ and $N_b$ respectively) are chosen using the relations $N_a=\frac{\phi_a}{\phi_{\rm tot}} N$ and $N_p=\frac{\phi_p}{\phi_{\rm tot}}  N$, where $\phi_a$ and $\phi_p$ are the fraction of active and passive particles that we choose for a particular simulation. To obtain a reliable estimate of the correlation length, we performed 50 independent simulations, each with a randomly chosen initial configuration. After each simulation reached a steady state in terms of both energy and correlation length, we obtained 50 snapshots of the system at regular intervals. In total, we obtained 2500 snapshots by averaging over the 50 simulations and the 50 snapshots from each simulation. By averaging over a large number of snapshots and simulations, we were able to obtain accurate estimates of the correlation length.

\section{System size effect}
\begin{figure}
    \centering
   \includegraphics[width=\linewidth]{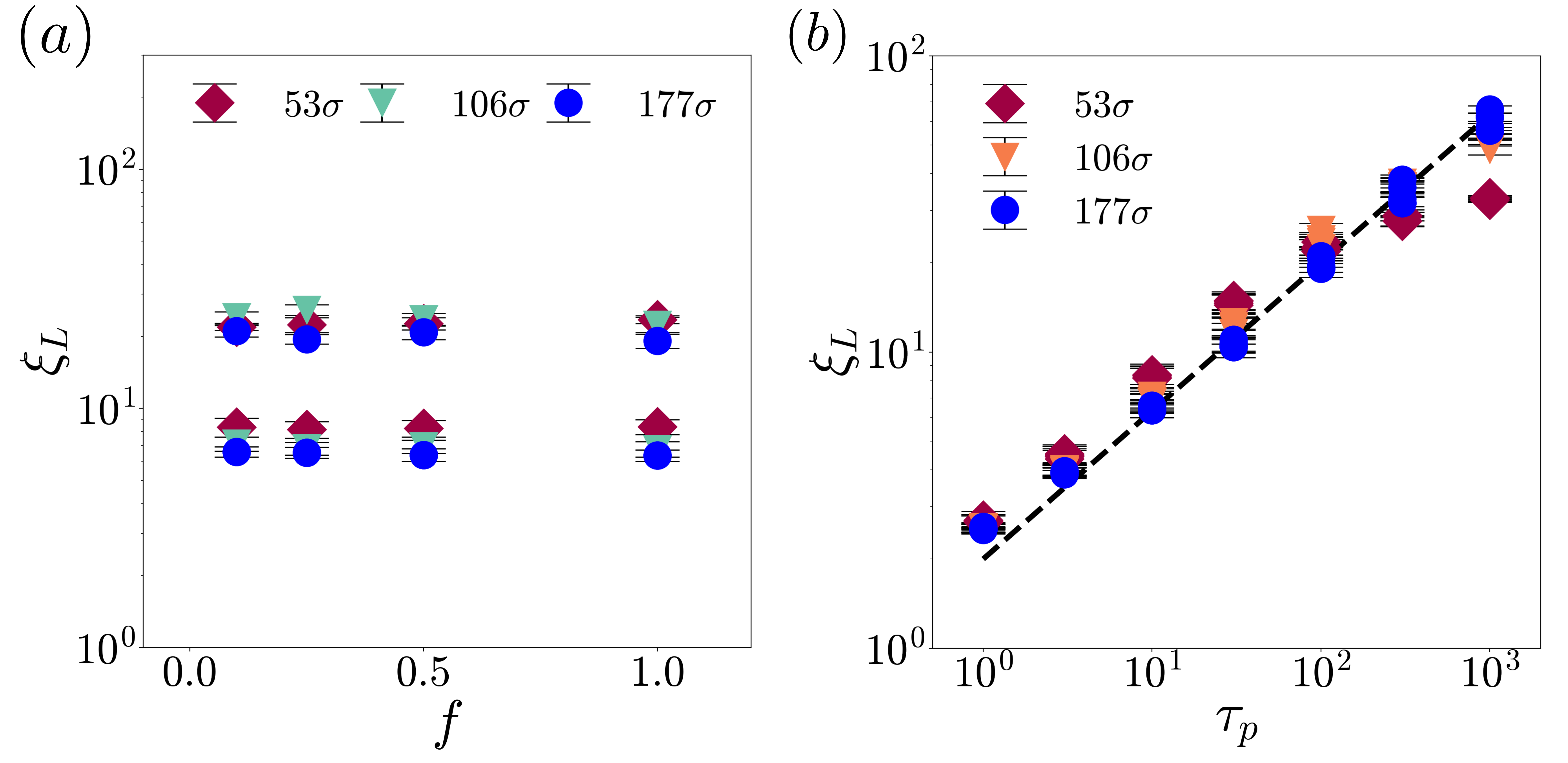}
    \caption{(a) Correlation length $\xi_L$ as a function of the magnitude of the active force for two different persistence times ($\tau_p=10$ and $100$) and three different system sizes (indicated by different colors and markers, see legend). The results indicate that the correlation length is unaffected by changes in the active force, for fixed values of persistence time and system size. (b) Correlation length $\xi_L$ as a function of persistence time $\tau_p$ for different values of the active force magnitude ($f=0.1$, $0.25$, $0.5$, $1$) and three different system sizes (indicated by different colors and markers). The results demonstrate that the correlation length exhibits a power-law growth (with exponent $\frac{1}{2}$) with increasing persistence time, and that this behavior is independent of the magnitude of the active force and system size.}
    \label{Fig-6}
\end{figure}

We conducted a systematic investigation of finite-size effects for a system with a fixed total density of $\phi_{\rm tot}=0.95$ ($\phi_a=0.01, \phi_p=0.94$), while using the same values for persistence time and self-propulsion force as in the plots  in the main text. Specifically, we explored three different system sizes ($L=53\sigma$, $106\sigma$, $177\sigma$) and found that the results exhibit a considerable degree of overlap across these sizes (Figure \ref{Fig-6}).

\section{Movies}
We have supplemented the information provided in the paper with a set of movies. 
\begin{itemize}
    \item Movie M1.mov displays a binary mixture of active and passive particles across different ($\phi_a$, $\phi_p$) values over time. As above, $\phi_a$ and $\phi_p$ denote the area fractions of active and passive particles, respectively. Both $\phi_a$ and $\phi_p$ values for each snapshot can be found in the corresponding row and column. The video showcases a significant correlation in velocity orientation when the overall density of the active and passive mixture is high enough. The self-propulsion force $f$ and persistence time $\tau_p$ are kept constant across the different densities, with values of 0.5 and 1000, respectively.
\end{itemize}
In the subsequent movies, we maintained a fixed area fraction of active particles at 0.01 while adding passive particles with varying area fractions to the system. The corresponding area fraction of passive particles for each movie is given below. In the visualization on the left hand side, blue particles represent active particles, while red particles represent passive particles. On the right hand side, the same system is shown with a color code that encodes the velocity orientation of the particles. In all of the movies, the self-propulsion force $f$ is set to 0.5, and the persistence time $\tau_p$ is 1000.
\begin{itemize}
    \item Movie M2.mov depicts a system with a passive particle area fraction of 0.15 and an active particle area fraction of 0.01 . In this case, the system remains in a homogeneous state, with active particles freely navigating the system. There is no apparent correlation in the velocity orientation of the particles in the system.
    
    \item Movie M3.mov displays a system with a passive particle area fraction of 0.5 and an active particle area fraction of 0.01. In this case, the excluded volume interaction between the particles becomes significant. Small domains with aligned velocity emerge in the system (on the right-hand side of the movie, small correlated domains with the same color can be observed). Notably, the active particles carve their own path through the passive medium by plowing through the particles (on the left-hand side of the movie).
    
    \item Movie M4.mov shows a system with a passive particle area fraction of 0.79 and an active particle area fraction of 0.01. It is evident that the average size of domains with aligned velocity increases, indicating that the active particles start to push the passive particles to move together coherently. This emphasizes the importance of a sufficiently large total area fraction of particles, so that strong velocity correlations can be observed.

    \item Movie M5.mov shows a system with a passive particle area fraction of 0.99 and an active particle area fraction of 0.01. Remarkably, it shows that the entire system becomes correlated, emphasizing the strong influence of active particles even at very low area fractions, particularly in systems with high total area fractions.  Note that in this case, a fraction of only $1\% $ 
    of active particles can correlate the entire system. The crystalline structure of the system is not essential here: similar results can be obtained in an amorphous systems consisting of a mixture of particle sizes.
    
\end{itemize}

\end{document}